\documentclass{aastex}
\usepackage{emulateapj5}

\newcommand{\mbf}[1]{\mbox{\boldmath $#1$}}
\newcommand{\mbfs}[1]{\mbox{\scriptsize\boldmath $#1$}}

\newcommand{\Eqn}[1]{Equation~(\ref{eqn:#1})}

\newcommand{\eqn}[1]{equation~(\ref{eqn:#1})}
\newcommand{\eqns}[1]{equations~(\ref{eqn:#1})}

\newcommand{\nbin}{65}
\newcommand{\nparam}{278}
\newcommand{\nfree}{25220}
\newcommand{\Ci}{\ensuremath{i}}

\newcommand{\trace}{{\rm tr}}

\newcommand{\Rotation}{{\bf R}}
\newcommand{\Boost}{{\bf B}}

\newcommand{\vRotation}[1][n]{\ensuremath{\Rotation_{\mbfs{\hat #1}}}}
\newcommand{\vBoost}[1][m]{\ensuremath{\Boost_{\mbfs{\hat #1}}}}

\newcommand{\rotat}{\ensuremath{\vRotation(\phi)}}
\newcommand{\boost}{\ensuremath{\vBoost(\beta)}}

\newcommand{\para}{\ensuremath{\Phi}}
\newcommand{\Rpara}[1][ ]{\ensuremath{\vRotation[v]{#1}(\para)}}

\newcommand{\selection}[1]{\ensuremath{\mbf{\delta}_{#1}}}

\newcommand{\pauli}[1]{\ensuremath{ {\mbf{\sigma}}_{#1} }}

\newcommand{\Smodelbase}[2]{\ensuremath{S^\prime_{k,m}({{\mbf x}#1}#2) }}

\newcommand{\Sfullmodel}[1][ ]{\Smodelbase{_{#1}}{;{\mbf{\eta}}}}

\newcommand{\modelbase}[2]{\ensuremath{{\mbf\rho}^\prime_m({{\mbf x}#1}#2) }}
\newcommand{\model}[1][ ]{\modelbase{_{#1}}{}}
\newcommand{\fullmodel}[1][ ]{\modelbase{_{#1}}{;{\mbf{\eta}}}}

\newcommand{\obs}{\ensuremath{ {\mbf\rho}^\prime_{m,n} }}
\newcommand{\Sobs}{\ensuremath{ S^\prime_{k,m,n} }}

\newcommand{\Sinv}[1][ ]{\ensuremath{ {S_{\rm inv}^{\prime\,{#1} }} }}

\newcommand{\psr}{PSR\,J0437$-$4715}

\shorttitle   {Polarimetric Calibration}
\shortauthors {W. van Straten}

\received{2003 December 5}
\begin{document}

\title{Radio Astronomical Polarimetry and Point-Source Calibration}

\author{W. van Straten}

\affil{Netherlands Foundation for Research in Astronomy}

\email{straten@astron.nl}

\begin{abstract}

A mathematical framework is presented for use in the experimental
determination of the polarimetric response of observatory
instrumentation.  Elementary principles of linear algebra are applied to
model the full matrix description of the polarization measurement
equation by least-squares estimation of non-linear, scalar
parameters. The formalism is applied to calibrate the center element
of the Parkes Multibeam receiver using observations of the millisecond
pulsar, \psr, and the radio galaxy, 3C 218 (Hydra~A).

\end{abstract}


\section {Introduction}

Polarization measurements provide additional insight into the
phenomena involved in both the emission and propagation of
electromagnetic radiation.  However, the processes of reception and
detection introduce instrumental artifacts that must be corrected
before meaningful interpretations of experimental data can be made.
Ideally, the instrumental response is estimated by observing at least
two calibrator sources of well-determined, polarized radiation
\citep{hbs96}.  Under the assumption that the observatory apparatus
respond linearly and remain stable between calibrator observations,
the known instrumental response is then inverted and used to calibrate
observations of other sources.

In the absence of a sufficient number of calibrator sources,
instrumental calibration may in some cases be performed by fitting the
available polarimetric data to a predictive model.  This model must
describe the observed sources of radiation, the response of the
instrument, and any additional propagation effects that arise in the
intervening media through which the signals are transmitted, such as
the Earth's ionosphere.  In order for the model to be constrained, at
least one component of the applied measurement equation must vary as
some function of one or more independent variables.  For example, the
constraining transformation may consist of the geometric projection of
the antenna receptors onto the plane of the sky, which may vary during
transit of the source for antennas without an equatorial mount.

For an altitude-azimuth mounted antenna, the receiver feed is rotated
about the line of sight through the parallactic angle, a constraint
that has been utilized in a number of previous studies, including
\cite{scr+84}, \cite{tfsw85}, \cite{xil91}, \cite{mck92}, and
\cite{joh02}.  In each of these cases, a matrix product is expanded to
yield a set of scalar equations that describe the measured Stokes
parameters as a function of parallactic angle; these are then solved
using conventional techniques.  This approach has two significant
limitations.  First, in order to simplify the derivation of the scalar
equations and their partial derivatives, it is necessary to make small
value approximations that are not generally valid.  Second, as any
alteration of the reception model necessitates a laborious expansion
of the matrix product, this approach is non-conducive to
experimentation with a variety of parameterizations.

In addition to these shortcomings, other fundamental limitations and
conceptual errors have been incorporated into previous treatments.
For instance, \citet{tfsw85} omit circular polarization from
consideration, an over-simplification that restricts further
application of their result to antenna with perfect, circularly
polarized receptors.  \citet{scr+84}, followed by \citet{xil91} and
\citet{mck92}, begin with the assumption that the complex gains of the
two polarizations may be independently calibrated prior to
determination of the cross-coupled antenna response.  However, this
approximation is accurate only when the differential gain and phase
transformations commute with the antenna response matrix, a condition
satisfied only when the feed receptors have well-determined,
orthogonal polarizations.  In general, the complex gains may be
accurately quantified only after the effect of the antenna feed is
included.

This error is essentially compounded in \citet{joh02} where, although
the instrumental response parameters are jointly estimated, the
measured data are first corrected using separately-determined,
inaccurate estimates of differential gain and phase.  In addition, the
observed Stokes parameters are incorrectly normalized by the total
intensity; a quantity such as the invariant interval must be
used \citep{scr+84,bri00}.  Finally, as only observations of an
unknown source are used to constrain the instrumental response
parameters, the solution derived by \citet{joh02} is not unique.
Rather, as shown in Appendix~\ref{app:degeneracy}, there exists a
degenerate set of solutions that can be resolved only with additional
observations of one or more sources with known circular polarization
and position angle.

In contrast to previous work, the treatment presented in this paper is
based entirely on a full matrix description of the polarization
measurement equation.  All of the matrix products, including those of
the required partial derivatives, are evaluated in software;
therefore, no small value approximations are necessary and a variety
of model specifications may be tested with relative ease.
Furthermore, all of the model parameters are jointly constrained using
observations of both unknown and partially known source polarizations.
Following a brief mathematical review in Section~2, the method
employed to solve the matrix equations is developed in Section~3.  Two
possible decompositions of the instrumental response are described in
Section~4, one of which is applied in Section~5 to the calibration of
the Multibeam receiver and downconversion system utilized by CPSR-II:
the 128\,MHz baseband recording and real-time processing system at the
Parkes Observatory.

\section {Polarized Radiation and Propagation}
\label{sec:review}

Electromagnetic radiation is described by the two complex-valued
components of the transverse electric field vector,
$\mbf{e}(t)=(e_0(t),e_1(t))$.  The measurable properties of
$\mbf{e}(t)$ are represented by the coherency matrix,
$\mbf{\rho}=\langle\mbf{e}(t)\mbf{\otimes\,e}^\dagger(t)\rangle$,
where the angular brackets denote time averaging and $\mbf{e}^\dagger$
is the Hermitian transpose of $\mbf{e}$.  The coherency matrix may be
written as a linear combination of Hermitian basis matrices
\citep{bri00},
\begin{equation}
{\mbf\rho} = {1\over2}\sum_{k=0}^3 S_k\pauli{k}
 = (S_0\,\pauli{0} + \mbf{S\cdot\sigma}) / 2,
\label{eqn:rho}
\end{equation}
where $\pauli{0}$ is the $2\times2$ identity matrix, $\mbf{\sigma} =
(\pauli{1},\pauli{2},\pauli{3})$ are the Pauli spin matrices, $S_0$ is
the total intensity, or Stokes~$I$, and $\mbf{S} = (S_1,S_2,S_3)$ is
the Stokes polarization vector.  As described in
Appendix~\ref{app:convention}, $\mbf{S} = (Q,U,V)$ in the Cartesian
reference frame applied in this paper.  The Stokes parameters may also
be expressed in terms of the coherency matrix \citep{ham00},
\begin{equation}
S_k = \trace(\pauli{k}\mbf{\rho}),
\label{eqn:stokes}
\end{equation}
where $\trace({\bf A})$ is the trace of the matrix, {\bf A}.

The propagation and reception of the electric field is represented by
the transformations of linear time-invariant systems.  Presented with
the input signal, $\mbf{e}(t)$, the output of a system is given by the
convolution, $\mbf{e}^\prime(t)={\bf j}(t)*\mbf{e}(t)$, where ${\bf
j}(t)$, is the complex-valued, $2\times2$ impulse response matrix.  By
the convolution theorem, this transformation is equivalent to
$\mbf{E^\prime}(\omega)={\bf J}(\omega)\mbf{E}(\omega)$, where {\bf J} is the
familiar Jones matrix.  When observed over a sufficiently narrow band,
the variation of {\bf J} with frequency is ignored, and polarimetric
transformations are represented by Jones matrix multiplications in the
time domain.  Under the operation, $\mbf{e}^\prime(t)={\bf
J}\mbf{e}(t)$, the coherency matrix is transformed as
\begin{equation}
{\mbf{\rho}^\prime}={\bf{J}}\mbf{\rho}{\bf{J}}^\dagger.
\label{eqn:congruence}
\end{equation}
This congruence transformation is called the polarization measurement
equation \citep{ham00}.  It forms the basis through which measured 
quantities are related to the intrinsic polarizations of the sources and 
used to model the unknown instrumental response.

\section {Maximum Likelihood Estimator}
\label{sec:merit}

To solve the polarization measurement equation using conventional
methods of least-squares minimization, it is necessary to design a
scalar figure-of-merit function and to calculate both its gradient and
curvature with respect to scalar model parameters.  Let $\mbf\eta$
represent the vector of scalar parameters that describe the model,
including the instrumental response and the polarizations of the
sources.  Furthermore, let $\mbf x$ represent the vector of
independent variables that constrain the measurement equation, such as
the observing frequency and epoch.  Given $\mbf x$, the model,
\begin{equation}
\fullmodel; \;\; 1\le m\le M,
\end{equation}
must predict the measured polarization of each source, where $M$ is
the number of sources.  In the interest of brevity, the model may also
be represented by $\model$.

Now consider $N_m$ independent observations of the $m^{\rm th}$
source, each made at a unique coordinate, ${\mbf x}_{m,n}$, to yield
the measured Stokes parameters and their estimated errors,
$\{S^\prime_k\pm\sigma_k\}_{m,n}$.  The best-fit model parameters will
minimize the objective merit function,
\begin{equation}
\chi^2(\mbf{\eta}) = \sum_{m=1}^M \sum_{n=1}^{N_m} \sum_{k=0}^3 
	{ [\Sobs - \Sfullmodel[m,n]]^2 \over \sigma_{k,m,n}^2 },
\label{eqn:merit}
\end{equation}
where $\Sfullmodel[m,n] = \trace[\pauli{k}\,\fullmodel[m,n]]$ are the
Stokes parameters of the $m^{\rm th}$ source as predicted by the model at
${\mbf x}_{m,n}$.  The gradient of $\chi^2$ with respect to the
scalar parameters, $\mbf\eta$, has components,
\begin{equation}
{\partial\chi^2\over\partial \eta_r} = -2 \sum_{m=1}^M \sum_{n=1}^{N_m}
\trace\left( {\mbf\Delta}_{m,n}({\mbf\eta}){\partial\fullmodel[m,n]\over\partial \eta_r} \right),
\end{equation}
where
\begin{equation}
{\mbf\Delta}_{m,n}({\mbf\eta}) = \sum_{k=0}^3 
	{ \Sobs - \Sfullmodel[m,n] \over \sigma_{k,m,n}^2 }\; \pauli{k}.
\end{equation}
Taking an additional partial derivative yields
\begin{equation}
{\partial^2\chi^2\over\partial \eta_s\partial \eta_r} = 
	-2 \sum_{m=1}^M \sum_{n=1}^{N_m} 
\trace\left( {\partial{\mbf\Delta}_{m,n}\over\partial \eta_s}
             {\partial\model[m,n]\over\partial \eta_r} \right),
\label{eqn:curvature}
\end{equation}
where, following the discussion in \S\,15.5 of Numerical Recipes
\citep{ptvf92}, the term containing a second derivative in
\eqn{curvature} has been eliminated.  Using \eqns{merit}
through~(\ref{eqn:curvature}), the Levenberg-Marquardt method is
applied to find the parameters that minimize $\chi^2(\mbf{\eta})$.

\section{Parameterization of the Model}
\label{sec:parameters}

It remains to specify the scalar values, $\mbf\eta$, that parameterize
the polarization measurement equation as well as the partial
derivatives of \fullmodel\ with respect to those parameters.
As in \eqn{rho}, the coherency matrix, ${\mbf\rho}_m$, of each input
source polarization is completely specified by the four Stokes
parameters, $S_{k,m}$.  The partial derivatives with respect to these
parameters are simply
\begin{equation}
{\partial{\mbf\rho}_m\over\partial S_{k,m}} = {\pauli{k}\over2}.
\end{equation}

Now, let the model of the instrumental response be represented by the
complex-valued $2\times2$ Jones matrix, ${\bf J}$.  If {\bf J}
satisfies \eqn{congruence} then it belongs to the set of solutions
given by ${\bf J}(\phi) = e^{\Ci\phi}{\bf J}$, where $\Ci=\sqrt{-1}$.
That is, the coherency matrix is insensitive to the absolute phase of
the signal and $\phi$ may be arbitrarily chosen, leaving seven degrees
of freedom with which to describe the instrumental response.  Two
possible parameterizations are considered: the algebraic decomposition
employed by \citet{ham00} and the phenomenological description of
\citet{bri00}.

\subsection{Algebraic Model}

Following \citet{ham00}, an arbitrary matrix, ${\bf J}$, is
represented by its polar decomposition,
\begin{equation}
{\bf J} = J \, \boost \, \rotat
\end{equation}
where $J=(\det{\bf J})^{1/2}$ and, as described in
Appendix~\ref{app:convention},
\begin{eqnarray}
\label{eqn:boost}
\boost &=& \cosh\beta\,\pauli{0} + \sinh\beta\,\mbf{\hat{m}\cdot\sigma}, \;
 {\rm and} \\
\rotat &=& \cos\phi\,\pauli{0} + \Ci\sin\phi\,\mbf{\hat{n}\cdot\sigma}.
\label{eqn:rotation}
\end{eqnarray}

The phase of the complex value, $J$, is set to zero and, as any
rotation about an arbitrary axis may be decomposed into a series of
rotations about three perpendicular axes, the instrumental response is
written as
\begin{equation}
{\bf J}_H = G \; \boost \prod_{k=1}^3 \vRotation[s_k](\phi_k),
\label{eqn:polar_decomposition}
\end{equation}
where $G=|J|$ is the absolute gain and $\mbf{\hat s_k}$ are the
orthonormal basis vectors defined in Appendix~\ref{app:convention}.
\Eqn{boost} may be written as \mbox{$\boost=b_0\,
\pauli{0}+\mbf{b\cdot\sigma}$}, where $b_0=\cosh\beta$ and
\mbox{$\mbf{b}=\sinh\beta\,\mbf{\hat m}$}. Noting that $b_0\ge1$ and
$b_0^2=1+\mbf{b\cdot b}$, the three degrees of freedom of $\boost$ are
specified by $\mbf{b}=(b_1,b_2,b_3)$.  Therefore, the instrumental
response is parameterized by $G$, $b_{1-3}$, and $\phi_{1-3}$.  The
partial derivatives of ${\bf J}_H$ with respect to these seven
parameters are calculated using products of
\begin{equation}
{\partial\boost\over\partial b_k} =
	{b_k\over\sqrt{1+\mbf{b\cdot b}}}\,\pauli{0} + \pauli{k},
\end{equation}
and
\begin{equation}
{\partial\vRotation[s_k](\phi_k)\over\partial \phi_k} = 
	-\sin\phi_k\,\pauli{0} + \Ci\cos\phi_k\,\pauli{k}.
\label{eqn:partial_basis_rotation}
\end{equation}

\subsection{Phenomenological Model}

Beginning with \citet{bri00}, the response of an ideal feed with two
orthogonally polarized receptors is given by
${\bf{S}}(\theta,\epsilon)=\vRotation[u](\epsilon)\vRotation[v](\theta)$.
Here, the receptors have ellipticities equal to $\epsilon$ and
mutually perpendicular orientations defined by $\theta$.  The basis
vectors, $\mbf{\hat q}$, $\mbf{\hat u}$, and $\mbf{\hat v}$, are
defined in Appendix~\ref{app:convention}.  Using this notation, a
receiver with non-orthogonal receptors is represented by
\begin{equation}
{\bf C} = \selection{0}{\bf S}(\theta_0,\epsilon_0)
	 + \selection{1}{\bf S}(\theta_1,\epsilon_1),
\label{eqn:receiver}
\end{equation}
where $\selection{a}$ is the $2\times2$ selection matrix,
\begin{equation}
\selection{a} = \left( \begin{array}{cc}
\delta_{0a} & 0 \\
0 & \delta_{1a} 
\end{array}\right),
\end{equation}
$\delta_{ab}$ is the Kronecker delta, and the product,
$\selection{a}{\bf B}$, returns a matrix that contains only the
$a^{\rm th}$ row of {\bf B}.  \Eqn{receiver} is equivalent to
equation~16 of \citet{bri00} and is used without making any first
order approximations or further assumptions about the feed.

The differential gain and phase of the instrument are represented by
$\vBoost[s_1](\gamma)$ and $\vRotation[s_1](\varphi)$ where, as in
\eqns{Boost} and~(\ref{eqn:Rotation}), $\gamma$ and $\varphi$
parameterize Lorentz boost and rotation transformations \citep{bri00}.
Including the absolute gain, $G$, the instrumental response is written
as
\begin{equation}
{\bf J}_B = G \; \vBoost[s_1](\gamma) \vRotation[s_1](\varphi) {\bf C}.
\label{eqn:physical}
\end{equation}
The partial derivatives of ${\bf J}_B$ with respect to its seven
scalar parameters, $G$, $\gamma$, $\varphi$, $\theta_{0-1}$, and
$\epsilon_{0-1}$, are calculated using \eqn{partial_basis_rotation}
and
\begin{equation}
{\partial\vBoost[s_k](\beta_k)\over\partial \beta_k} = 
	\sinh\beta_k\,\pauli{0} + \cosh\beta_k\,\pauli{k}.
\label{eqn:partial_basis_boost}
\end{equation}

\section{Application}
\label{sec:application}

Radio pulsar observations provide an excellent source of data with
which to constrain the polarization measurement equation. The often
highly polarized state of a pulsar signal can vary significantly as a
function of pulse longitude.  When integrated over a sufficient number
of spin periods, mean polarimetric pulse profiles generally exhibit
excellent stability on timescales much longer than those over which
calibration observations are made.  Consequently, multiple on-pulse
longitudes from a single pulsar may be included as unique and stable
input source polarizations, greatly increasing the number of available
constraints when compared with non-pulsed sources.  Furthermore, any
non-pulsed background polarization is effectively eliminated by
subtracting the off-pulse mean from each integrated pulse profile.

The millisecond pulsar, \psr, represents an ideal candidate for use in
the regular calibration of the apparatus at the Parkes Observatory. As
part of the high-precision pulsar timing program, it is often observed
from rise to set, providing measurements with a wide range in
parallactic angle.  The model that describes these observations has
the form,
\begin{equation}
{\mbf\rho}^\prime_m(\para)
	= {\bf J}\Rpara\mbf{\rho}_m\Rpara[^\dagger]{\bf J}^\dagger,
\label{eqn:model}
\end{equation}
where $\Rpara$ is the rotation about the line of sight by the
parallactic angle, \para.

As demonstrated in Appendix~\ref{app:degeneracy}, additional
calibrator observations are required in order to uniquely determine
the solution to \eqn{model}.  The Parkes Multibeam receiver is
equipped with a noise diode that ideally injects a 100\% linearly
polarized signal with a position angle of 45 degrees into the receiver
feed horn.  This reference source is switched using a signal generator
and the observed square waveform is integrated modulo its period.
Ideally, the on-pulse longitudes of the input reference signal contain
additional flux with Stokes parameters given by $C_0[1,0,1,0]$, where
$C_0$ is the reference flux density.  Under this assumption, inclusion
of reference signal observations breaks the degeneracy by constraining
the boost, $\beta_v$, which mixes Stokes~$I$ and $V$, and the
rotation, $\phi_v$, which mixes Stokes~$Q$ and $U$.  In a separate
flux calibration procedure, the reference signal is observed
simultaneously with the bright Fanaroff-Riley type I radio galaxy, 3C
218 (Hydra~A), producing absolute flux estimates of both the system
temperature and reference flux density, $C_0$.

\subsection{Observations}
\label{sec:observation}

Dual-polarization observations of \psr\ and Hydra~A were made on 19
and 20 July 2003 using the center element of the Parkes Multibeam
receiver.  Two 64\,MHz bands, centered at 1341 and 1405\,MHz, were
two-bit sampled and processed by CPSR-II, the second generation of the
Caltech-Parkes-Swinburne Recorder.  In order to maintain optimal
linear response during the digitization process, the detected power is
monitored and the sampling thresholds are updated approximately every
30 seconds.  In addition, the baseband data reduction software
corrects quantization distortions to the voltage waveform using the
dynamic level-setting technique \citep{ja98}.  Phase-coherent
dispersion removal is performed while synthesizing a 128-channel
filterbank \citep{jcpu97}; the Stokes parameters are then detected and
integrated as a function of topocentric pulse phase.  Data are
averaged for five minute intervals, producing uncalibrated mean pulse
profiles with 2048 phase bins, or an equivalent time resolution of
approximately 2.8\,$\mu$s.

The observed flux density varies significantly between pulsar and
Hydra~A observations, resulting in large changes to the digitization
sampling thresholds.  This difference is best modeled using the
phenomenological decomposition of \eqn{physical}, applying separate
absolute gain, differential gain, and differential phase terms to the
two sets of observations.  Also, in order to account for phase drifts
on timescales of a few hours, the differential phase is modeled to
vary as a cubic polynomial function of time.  The three signal path
transformations corresponding to the three observed sources are
summarized in Table~\ref{tab:model}.

\begin{figure*}
\centerline{\includegraphics[angle=-90,width=86mm]{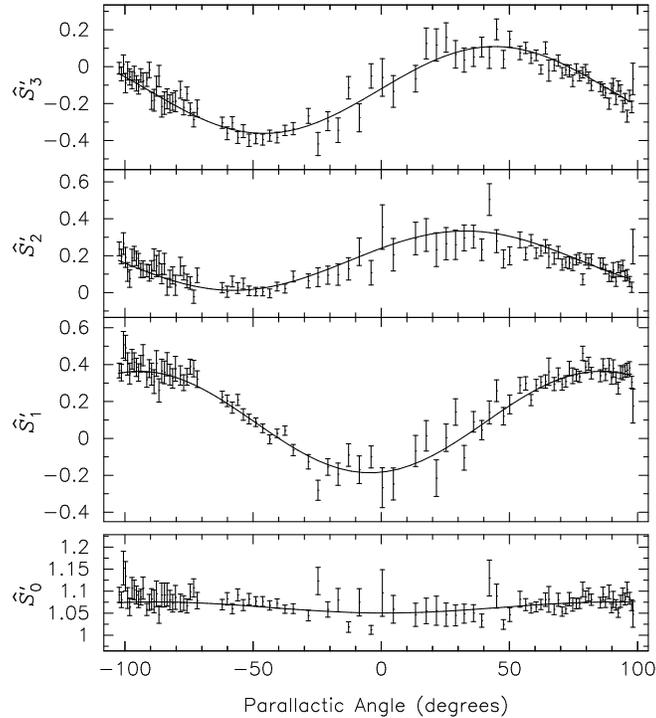}}
\caption {Observed Stokes parameters from \psr, normalized by the invariant
interval and plotted as a function of parallactic angle.  These data
correspond to a pulse phase of approximately 0.485 in
Figure~\ref{fig:psr}, as observed in a single 500\,kHz channel
centered at 1324.75\,MHz.  The 
Stokes parameters predicted by the best-fit model
are drawn with solid lines.}
\label{fig:stokes}
\end{figure*}

The flux of the pulsar also varies between observations, a result of
both intrinsic intensity fluctuations and interstellar scintillation.
Therefore, the measured Stokes parameters are normalized by the
invariant interval,
\begin{equation}
\Sinv = 2(\det\obs)^{1/2} = (S_0^{\prime\,2} - S^{\prime\,2})^{1/2},
\end{equation}
where the polarized intensity, $S^\prime = |\mbf{S}^\prime|$.  In
order to avoid division by small or negative values, those data points
with $\Sinv<\sigma_0$ are discarded.  Assuming that the measurement
errors in each of the Stokes parameters are independent of each other,
the estimated errors in the normalized Stokes parameters are given by
\begin{eqnarray}
\label{eqn:normalized_Ierror}
{\hat\sigma}_0^2 & = &  \Sinv[-6]\, S^{\prime\,2}\,
	  ( S^{\prime\,2}\sigma_0^2 + S_0^{\prime\, 2}\sigma_s^2 ), \;
 {\rm and} \\
\label{eqn:normalized_perror}
{\hat\sigma}_k^2 & = &  \Sinv[-4]\,
	[ ( \Sinv[2] + 2S_k^{\prime\,2} )\,\sigma_k^2 + S_k^{\prime\,2}\sigma_{\rm inv}^2 ],
\end{eqnarray}
where $1\le k \le 3$,
\begin{equation}
S^{\prime\,2}\sigma_s^2 = \sum_{k=1}^3 S_k^{\prime\,2}\sigma_k^2,
\end{equation}
and $\Sinv[2]\sigma_{\rm inv}^2 = S_0^{\prime\,2}\sigma_0^2 +
S^{\prime\,2}\sigma_s^2$.  Note that $\hat\sigma_0$ is approximately
proportional to $S^\prime$ when $S^\prime \ll S_0$.  Consequently, the
normalized total intensity is plotted on a separate scale in
Figure~\ref{fig:stokes}, in which are displayed the normalized Stokes
parameters from one of the \nbin\ pulse longitudes used to constrain
the model.

\subsection{Results}
\label{sec:results}

\begin{figure*}
\centerline{\includegraphics[angle=-90,width=86mm]{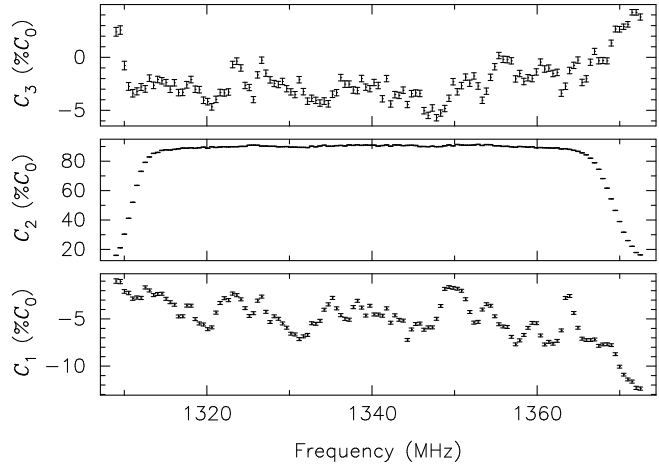}}
\caption {Stokes parameters of the noise diode reference signal,
plotted as a function of observing frequency.  The modeled values of
Stokes~$Q$, $U$, and~$V$ ($C_1$, $C_2$, and $C_3$) are specified as
percentages of the reference flux density, $C_0$.  Error bars correspond
to the formal standard deviations of the model parameters derived from
the best-fit covariance matrix. }
\label{fig:cal}
\end{figure*}

Initial results indicated that the reference signal produced by the
noise diode of the Parkes Multibeam receiver is not actually 100\%
linearly polarized.  Rather, as shown in Figure~\ref{fig:cal}, the
reference signal consists of $\sim 90\%$ linear and $3\%$ circular
polarization, its position angle is not exactly 45 degrees
(i.e. Stokes~$Q \ne 0$), and it is severely depolarized at the edges
of the band due to frequency aliasing during downconversion.
Consequently, the reference signal cannot be trusted as a source with
known position angle and degree of circular polarization.  Also,
approximately 7\,MHz from both the top and bottom of the band must be
discarded as irreversibly corrupted.

In order to constrain $\beta_v$, it is noted that Hydra~A has less
than 0.1\% circular polarization \citep{rcm+75}.  Therefore, the
off-pulse longitudes of the flux calibration observations serve as a
source of negligible circular polarization in the remainder of this
development.  The unknown rotation about the line of sight, $\phi_v$,
is artificially constrained by assuming that $\theta_0 = 0$;
consequently, all position angles are measured with respect to the
orientation of receptor~0.  In modeling the reference signal,
Stokes~$I$ is set to unity, producing an intermediate flux scale in
units of the reference flux density; Stokes~$Q$, $U$, and $V$ are
varied as free model parameters.

The reception model is solved independently for each of the 256
frequency channels in the two 64\,MHz bands.  In each channel are a
total of \nparam\ free model parameters, corresponding to the Stokes
parameters of the \nbin\ selected pulse longitudes, the 12
instrumental parameters listed in Table~\ref{tab:model}, and the 6
unknown Stokes parameters shared between the noise diode and Hydra~A.
These parameters are constrained by approximately \nfree\ measured
values, primarily derived from the selected pulse longitudes of the 97
pulsar observations.  Indicating a good fit in each frequency channel,
the merit function is on average approximately equal to the number of
degrees of freedom, $\nu$, such that
\mbox{$\langle\chi^2/\nu\rangle\sim1.05$}.  Under the assumption that
the errors in the measured Stokes parameters are normally distributed,
the curvature matrix of \eqn{curvature} is inverted to yield the
covariance matrix of the standard errors in each of the model
parameters.

\begin{figure*}
\centerline{\includegraphics[angle=-90,width=86mm]{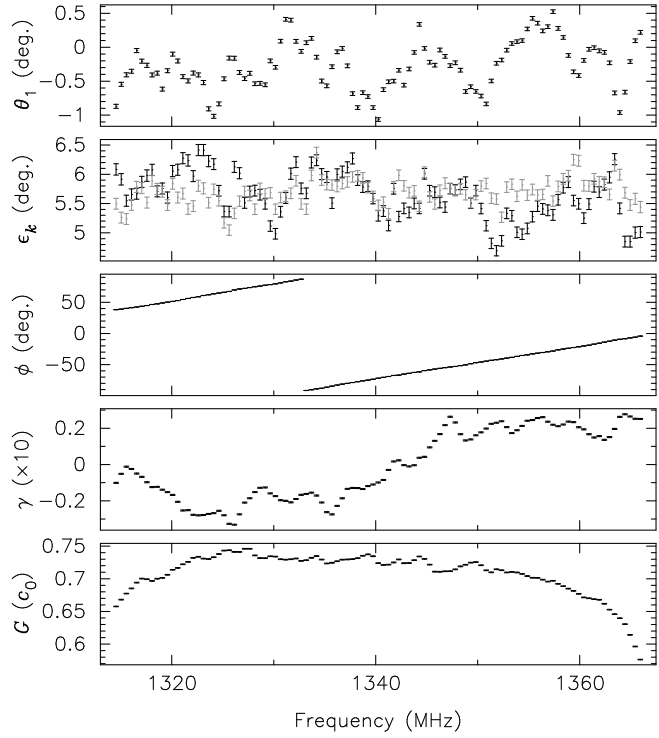}}
\caption {Best-fit model parameters as a function of observing
frequency.  From top to bottom are plotted the orientation of
receptor~1 with respect to receptor~0, $\theta_1$, the ellipticities
of the receptors, $\epsilon_k$, the differential phase,
$\phi=\varphi(0)$, the differential gain, $\gamma$, and the absolute
gain, $G$, specified in units of the square root of the reference flux
density, $c_0=\sqrt{C_0}$.  In the panel showing the ellipticities,
black and grey correspond to receptors~0 and~1, respectively. As in
Figure~\ref{fig:cal}, error bars correspond to the formal standard
errors of the model parameters.}
\label{fig:solution}
\end{figure*}

In Figure~\ref{fig:solution} are plotted the instrumental response
parameters from one of the CPSR-II bands at the reference epoch.  The
linear dependence between differential phase and frequency indicates a
signal path length mismatch between the two polarizations.  The
receptor ellipticities have an average value of $\sim5.7$ degrees,
corresponding to a rotation of the polarization vector about the
$U$-axis by $\sim0.2$ radians.  Consequently, the degree of mixing
between linear and circular polarizations reaches approximately $20$\%
for uncalibrated signals with polarization vectors lying near the
$Q-V$ plane.  Clearly, this level of distortion cannot be treated as a
second-order effect.

\begin{figure*}
\centerline{\includegraphics[angle=-90,width=86mm]{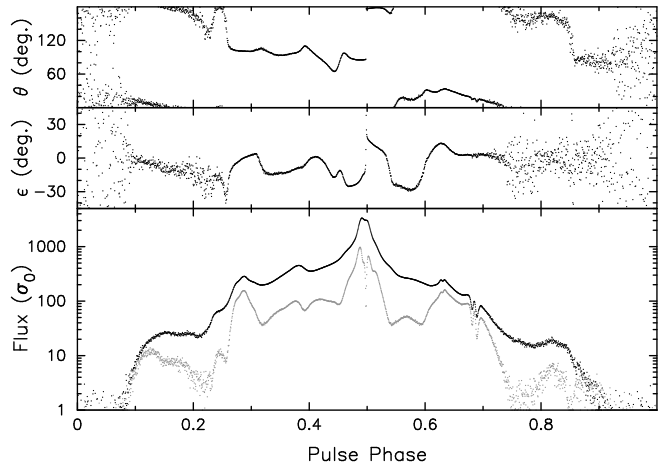}}
\caption {Mean polarization of \psr, plotted as a function of pulse
phase using polar coordinates: orientation, $\theta$, ellipticity,
$\epsilon$, and polarized intensity, $S=|\mbf{S}|$ (plotted in grey
below the total intensity, $S_0$).  Flux densities are normalized by
$\sigma_0$, the r.m.s. of the off-pulse total intensity phase bins.
Data were integrated over a 64\,MHz band centered at 1341\,MHz for
approximately 8~hours.  }
\label{fig:psr}
\end{figure*}

The mean polarimetric pulse profile of \psr\ is plotted in
Figure~\ref{fig:psr}.  Here, the position angle is equal to $\theta$
and the colatitude is given by $\pi/2 - 2\epsilon$.  When compared
with cylindrical coordinates, as plotted in Figure~3 of \cite{van02},
spherical coordinates offer a number of conceptual advantages.  For
instance, both the total and polarized intensities may be plotted on
the same logarithmic scale, thereby enhancing the smaller features of
these profiles.  In addition, the chosen normalization allows the
statistical significance of the measured quantities to be approximated
without explicitly plotting the estimated error bars.  Finally, the
correlations between rapid transitions in position angle and dips in
polarized intensity are more obviously apparent, for example, at pulse
phases approximately equal to 0.26, 0.50, and 0.84.  These regions are
interpreted as transitions between two highly polarized, nearly
orthogonal, superposed modes of radiation, possibly the natural modes
of the pulsar magnetospheric plasma \citep[and references
therein]{pet01}.

\section{Conclusions}
\label{sec:conclusions}

A fundamentally different approach to modeling the reception and
propagation of polarized radiation has been presented in this paper.
Based on the formalism developed in Sections~3 and 4, a detailed
calibration model has been constructed from simple, modular
components, incorporating multiple signal paths and source
polarizations as constraints.  Many of the limiting assumptions made
in previous treatments have been eliminated in this analysis, enabling
its application in a wider variety of experiments.  This development
is increasingly relevant in the context of design considerations for
the next generation of radio telescopes and instrumentation.  For
instance, it has been proposed that the Square Kilometer Array (SKA)
will be an interferometric array of a large number of low-cost
antennas.  Especially in the case of fixed dipole pairs, the
polarizations of the receptors may be highly non-orthogonal;
therefore, it is important to develop more sophisticated theories and
techniques of polarimetric calibration.

Due to the current lack of a standard catalog of multi-frequency
polarimetric pulse profiles, it is difficult to confirm the validity
of any calibration technique.  Future efforts should include the
establishment of a set of well-calibrated sources, including a number
of stable pulsars that are monitored on a regular basis at more than
one observatory and at a variety of radio frequencies.  Especially at
lower frequencies, other propagation effects such as those arising in
the ionosphere will have to be included in the reception model applied
to these observations.  Once established, a relatively short
integration on a bright, calibrated pulsar could be used to quickly
determine the instrumental response without the need for
time-intensive techniques such as the one described in this paper.

In order that the results presented in Section~\ref{sec:results} may
be used to calibrate other observations, they are stored as binary
table extensions of a PSRFITS file \citep{hsm04}, a pulsar data format
defined using the Flexible Image Transport System \citep{hfg+01}.
Both the PSRFITS file format and its associated visualization and
reduction software have been openly developed in an effort to
facilitate the exchange of pulsar astronomical data between
observatories and research groups.

As the current work is based on the Jones calculus, it is limited to
describing only ``non-depolarizing'' or ``pure'' component
transformations; it is not possible to represent the conversion of a
fully polarized signal into a partially polarized one using Jones
operators \citep{hbs96}.  This is demonstrated by noting that det({\bf
AB}) = det({\bf A})det({\bf B}) and, for a fully polarized signal,
$\det(\mbf{\rho})=0$.  Certain components of the instrumental
apparatus may not be pure and therefore may require a Mueller matrix
in order to model their effect.  For example, the process of two-bit
quantization may act to depolarize strong input signals, an effect
that has yet to be studied in rigorous detail.  Other phenomenon, such
as bandwidth depolarization, may not require a Mueller matrix
description if they can be treated by the application of
phase-coherent matrix convolution \citep{van02}.

The formalism developed in this paper may be equally well applied to
the polarimetric calibration of a phased array.  In this case, it is
not necessary to describe the complete instrument in terms of the
complex gains, orientations, and ellipticities of its individual
receptors.  Rather, it is more useful to consider only the total
instrumental transformation given by the sum of the Jones matrices of
the individual antennas.  This development will be the subject of
future work.

\acknowledgements

The Parkes Observatory is part of the Australia Telescope which is
funded by the Commonwealth of Australia for operation as a National
Facility managed by CSIRO.  This research greatly benefited from
lectures presented by Johan Hamaker from 14 January to 18 March, 2003.
Thanks also to the Swinburne University of Technology Pulsar Group for
providing the CPSR-II observations and to Ben Stappers for advice on
the text.

\begin{table*}
\caption{Signal Path Transformations}
\begin{center}
\begin{tabular}{c|c}
\tableline
\tableline
Source & Transformation \\
\tableline
Noise Diode & ${\bf J}(t)=G\;\vBoost[s_1](\gamma)\vRotation[s_1](\varphi(t)){\bf C}$ \\
\psr        & ${\bf J}(t)\Rpara$ \\
3C 218      & $G_H\vBoost[s_1](\gamma_H)\vRotation[s_1](\varphi_H){\bf C}$ \\
\tableline
\end{tabular}
\end{center}
\label{tab:model}
\end{table*}

\appendix

\section{Geometric Interpretation}
\label{app:convention}

The $2\times2$ complex matrices with unit determinant may be
parameterized by
$\exp[(\beta\,\mbf{\hat{m}}+\Ci\phi\,\mbf{\hat{n}})\mbf{\cdot\sigma}]$.
The unit vectors, $\mbf{\hat{m}}$ and $\mbf{\hat{n}}$, as well as the
dimensionless values, $\beta$ and $\phi$, have meaningful geometric
interpretations in the four-dimensional space of the Stokes
parameters.  Under the congruence transformation of \eqn{congruence},
the Hermitian matrices,
\begin{equation}
\label{eqn:Boost}
\boost = \exp (\beta\,\mbf{\hat{m}\cdot\sigma})
       = \cosh\beta\,\pauli{0} + \sinh\beta\,\mbf{\hat{m}\cdot\sigma},
\end{equation}
effect a Lorentz boost of the Stokes 4-vector along the axis
$\mbf{\hat m}$ by an impact parameter $2\beta$.  Likewise, the unitary
matrices,
\begin{equation}
\label{eqn:Rotation}
\rotat = \exp (\Ci\phi\,\mbf{\hat{n}\cdot\sigma})
       = \cos\phi\,\pauli{0} + \Ci\sin\phi\,\mbf{\hat{n}\cdot\sigma},
\end{equation}
rotate the Stokes polarization vector about the axis $\mbf{\hat n}$ by
an angle $2\phi$.  In applying this geometric representation,
the following conventions are used.
First, the Pauli spin matrices are defined as
\begin{eqnarray}
\pauli{1} = \left( \begin{array}{cc}
1 & 0 \\
0 & -1 
\end{array}\right)
&
\pauli{2} = \left( \begin{array}{cc}
0 & \;1 \\
1 & \;0 
\end{array}\right)
& 
\pauli{3} = \left( \begin{array}{cc}
0 & -i \\
i & 0
\end{array}\right).
\label{eqn:pauli}
\end{eqnarray}
Second, the three-dimensional space of the Stokes polarization vector,
$\mbf{S} = (S_1,S_2,S_3)$, is spanned by the orthonormal basis
vectors, $\mbf{\hat s_k}$, such that $\mbf{{\hat s_k}\cdot S}=S_k$ and
$\mbf{{\hat s_k}\cdot\sigma}=\pauli{k}$.  Finally, Stokes~$Q$, $U$,
and $V$ are calculated by the projection of $\mbf{S}$ onto the Stokes
unit vectors, $\mbf{\hat q}$, $\mbf{\hat u}$, and $\mbf{\hat v}$,
respectively. This relationship is summarized by $\mbf{p}=(Q,U,V)={\bf
R}^T\mbf{S}$, where ${\bf R}=(\mbf{\hat q} \; \mbf{\hat u} \;
\mbf{\hat v})$ is a three-dimensional rotation matrix with columns
defined by the Stokes unit vectors.  The orientation of these basis
vectors with respect to $\mbf{\hat s_k}$ depends upon the reference
frame in which the electric field vector is represented.  In the
Cartesian basis, the plane wave propagates along the $z$-axis, the
electric field is decomposed into its projection along the $x$ and $y$
axes, and ${\bf R}$ is equal to the $3\times3$ identity matrix.

\section{Degeneracy under Commutation}
\label{app:degeneracy}

The Pauli spin matrices satisfy
$\pauli{i}\,\pauli{j} = \delta_{ij}\,\pauli{0} + i\epsilon_{ijk}\,\pauli{k}$,
where $\epsilon_{ijk}$ is the permutation symbol and summation over
the index $k$ is implied.  Using this identity, the product of two
arbitrary matrices, {\bf A} and {\bf B}, may be expanded to yield
\begin{equation}
{\bf AB} = 
 (a\,\pauli{0}+\mbf{a\cdot\sigma})(b\,\pauli{0} + \mbf{b\cdot\sigma})
         = 
 (ab + \mbf{a\cdot b})\pauli{0} + (a\mbf{b} + b\mbf{a} 
    + \Ci\,\mbf{a\times b})\mbf{\,\cdot\sigma}.
\label{eqn:mult}
\end{equation}
In this form, it can be seen that {\bf A} and {\bf B} commute if and
only if $\mbf{a}$ and $\mbf{b}$ are collinear, enabling simple
statements to be made regarding the uniqueness of solutions to the
polarization measurement equation.

For example, consider the parallactic angle rotation given by
$\Rpara=\cos\para\,\pauli{0}+\Ci\sin\para\,\mbf{\hat{v}\cdot\sigma}$.
As illustrated by \eqn{mult}, any matrix of the form, ${\bf V}=
V_0\,\pauli{0} + V\mbf{\hat{v}\cdot\sigma}$, commutes freely with
\Rpara.  Therefore, if both {\bf J} and ${\mbf\rho}_m$ satisfy
\eqn{model}, then ${\bf J}^v = {\bf JV}^{-1}$ and ${\mbf\rho}_m^v =
{\bf V}{\mbf\rho}_m{\bf V}^{\dagger}$ are also solutions to this
equation.  This degeneracy exists regardless of the parameterization
of {\bf J} or ${\mbf\rho}_m$, proving that no unique solution to the
polarization measurement equation may be derived solely from
observations of unknown sources at multiple parallactic angles.

The consequences of this degeneracy are illustrated by noting that
{\bf V} is unimodular and may be decomposed into a boost along the
$\mbf{\hat{v}}$ axis, $\vBoost[v](\beta_v)$, and a rotation about this
axis, $\vRotation[v](\phi_v)$.  That is, unless {\bf V} is determined
through some other means, there remains an unknown position angle
error as well as an unknown degree of mixing between the total
intensity and circular polarization.  An observation of a (possibly
unpolarized) source with a well-determined degree of circular
polarization may be used to determine $\beta_v$; additionally, an
observation of a linearly polarized source with a known position angle
(such as the noise diode included in many receiver packages) may be
used to constrain $\phi_v$.


\end{document}